\documentclass[reprint,showpacs,showkeys,aps,prb]{revtex4-2}
\usepackage[T1]{fontenc}
\usepackage[utf8]{inputenc}
\usepackage[english]{babel}
\usepackage{graphicx}
\usepackage{times}
\usepackage{mathptmx}
\usepackage{eurosym}
\usepackage{amssymb,amsmath,amsfonts}
\usepackage{bm}
\usepackage{latexsym}
\usepackage{subfigure}
\usepackage{url}
\usepackage{hyperref}
\usepackage{tensor}
\usepackage{xfrac}
\usepackage{multirow}
\usepackage{tabulary}
\usepackage{fancyhdr}
\usepackage{MnSymbol}
\usepackage{float}
\usepackage{booktabs}
\usepackage[flushleft]{threeparttable}
\usepackage{lineno}
\usepackage{balance}
\usepackage{float}
\usepackage[usenames]{color}
\usepackage{lipsum}

\begin{document}

\title{Chemical potential influence on the condensation energy\break from a Boson-Fermion model of superconductivity}
	\author{I. Chávez$^{\star}$}
	\author{P. Salas}
	\author{O.A. Rodríguez}
	\author{M.A. Solís}
	\affiliation{Instituto de Física, Universidad Nacional Autónoma de México,\break
Apdo. Postal 20-364, 01000 Mexico City, MEXICO}
	\author{M. de Llano}
	\affiliation{Instituto de Investigaciones en Materiales, Universidad Nacional Autónoma de México,\break 04510 Mexico City, MEXICO}
	


\date{\today}

\begin{abstract}
Influence of the temperature dependent chemical potential on the condensation energy from a ternary Boson-Fermion model of superconductivity is reported, it consist of unbound electrons/holes which are fermions plus two-electron and two-hole Cooper pairs which are bosons. When solving simultaneously the set of equations of the mixture (two gap-like equations, one for electron pairs and another one for hole pairs, plus the particle number conservation equation) within the weak-coupling (BCS regime), the resulting  superconducting chemical potential shows a shift from its normal state counterpart, which is related to both the magnitude of the temperature-dependent superconducting gap and to the Fermi energy of the superconductor. As predicted by van der Marel in the early 1990s we also find that the superconducting chemical potential has a prominent kink at critical temperature $T_c$, which in turn coincides with the normal state chemical potential. Also there is discontinuity in its first derivative which directly affects the magnitude in the specific heat jump. We show that the difference between the superconducting and normal state chemical potentials is of the same order of magnitude as the corresponding difference between the thermodynamic potentials for the mixture, and must therefore be accounted for in the condensation energy calculations instead of ignoring it as is done often. The condensation energy obtained here shows very good agreement with experimental data for elemental superconductors.
\\[0.25cm]
$\star$ Corresponding author; email: israelito@fisica.unam.mx
\end{abstract}


\maketitle

\section{Introduction}

Since the BCS \cite{BCS} theory two recurring approximations have been made, one is that the chemical potential is equal to the Fermi energy from zero temperature up to the transition temperature $T_c$. This is complemented with the assumption that the density of states (DOS) around the Fermi level is a constant. Both approximations lead to solve only the energy-gap equation, whose gap magnitude is much higher than the magnitude of the difference between the normal and superconducting chemical potentials; this justifies taking them as equal. However, this difference is of the same magnitude as the difference between the normal and superconducting thermodynamic potentials, so it should not be ignored in the condensation energy calculations.

To obtain the energy gap and the chemical potential both as functions of temperature, one needs to add the equation of the number of system particles to be solved simultaneously. This idea was suggested for the first time by Schrieffer \cite{schrieffer63} (1963) and later developed by Keldysh \cite{keldysh} (1965), Popov \cite{popov} (1966), Labb\'{e} \textit{et al}. \cite{friedel} (1967). In 1969 Eagles \cite{eagles} solved this set of equations. Thus many authors developed the so-called crossover picture, like Miyake \cite{miyake} (1983), Nozi\`{e}res \cite{nozieres} (1985), Ranninger \textit{et al.} \cite{ranninger} (1988), Randeria \textit{et al}. \cite{randeria} (1989), van der Marel \cite{vandermarel} (1990), Bar-Yam \cite{baryam} (1991), Drechsler and Zwerger \cite{drechsler} (1992), Haussmann \cite{haussman} (1993), Pistolesi and Strinati \cite{pistolesi} (1996), among others. In this crossover picture Cooper pairs \cite{cooper} can be considered to undergo \cite{dellano2006} a Bose-Einstein condensation (BEC).
\par
As a function of the pair interaction magnitude, it is expected that the chemical potential in the weak coupling (BCS) regime must behave nearly as that of an ideal Fermi gas (IFG), while in the strong coupling extreme it must behave as the chemical potential of an ideal Bose gas (IBG). This picture describes well the behavior of $\mu/E_F$ vs. $1/k_F\,a_s$ as seen in Fig. 3 of Ref. \cite{carter95} for $T=0$, with $k_F$ the Fermi wave number and $a_s$ the s-wave scattering length of two-body interactions. If $1/k_F\,a_s \to -\infty$, $\mu(0)/E_F \to 1$ (BCS extreme) while $1/k_F\,a_s \to +\infty$, $\mu(0)/E_F \to -\infty$ as the coupling increases (Bose extreme). Here we address a ternary Boson-Fermion (BF) model of superconductivity consisting of unbound electrons/holes which are fermions, plus two-electron Cooper pairs (2eCP) and \textit{explicitly} two-hole Cooper pairs (2hCP) both as bosons. This BF model \cite{GBEC1, GBEC2, GBEC3, GBEC4} incorporates all the previously mentioned ideas, so one is able to recover the BCS theory and the BCS-Bose crossover \cite{chavez17a, chavez17b}, among other results. However, here we focus on the weak-coupling extreme.
 
It is worth mentioning that in the early 1990s van der Marel \cite{vandermarel} solved the BCS coupled equations at a finite temperature and showed that the superconducting chemical potential had a ``kink'' at $T_c$ and that at $T=0$ it is related to the energy gap as
\begin{equation}
\mu(0) - E_F = -\frac{\Delta(0)^{2}}{2E_F}
\label{vandermarel}
\end{equation}
when $\hbar\omega_D >> \Delta$. This result is the same as that obtained by Randeria \textit{et al}. \cite{randeria} in the 2D case. The kink in the chemical potential $\mu(T)$ and the jump in its derivative are directly related to the jump magnitude in the specific heat at $T_c$. This discontinuity precisely at $T_c$ of the derivative of the temperature-dependent chemical potential is also seen in Ref.~\cite{vandermarel}.

Our interest here is to calculate the temperature dependent condensation energy, from the difference between the superconducting and normal Helmholtz free energy without ignoring the small difference between the corresponding chemical potentials as most authors have done, e.g., Fetter \& Walecka (FW) \cite{fetter}. To do this, we obtain the energy gap and chemical potential, both as functions of temperature and free of approximations, within the BF model of superconductivity \cite{GBEC1, GBEC2, GBEC3, GBEC4}.
 
We find that the difference between the superconducting chemical potential for $0<T<T_c$ and the normal chemical potential at $T = 0$ is different from zero and the same order of magnitude as that of the condensation energy, which in turn differs from Ref.~\cite{fetter} where $\mu_s = \mu_n$ is taken. Furthermore, we do not assume that the density of states remains constant. We get the temperature dependent energy gap and compare our results with Eq.~\eqref{vandermarel} not only for $T = 0$. We also find the kink in the chemical potential in the weak coupling (BCS) extreme and study its influence on the Helmholtz free energy and therefore also on the condensation energy.

In Sec. II we recall the BF Hamiltonian and the main equations of a generalized Bose-Einstein condensation (GBEC) formalism \cite{GBEC1, GBEC2}. The BF Hamiltonian \cite{GBEC3} was diagonalized to find the grand thermodynamic potential and the chemical potential as functions of temperature, the number density of two-electron Cooper pairs (2eCPs) and the number density of two-hole Cooper pairs (2hCPs); in Sec. III we solve the three main equations of this ternary BF gas and take the special case of 50\%-50\% proportions between 2eCPs and 2hCPs, which recovers the BCS theory. With the superconductor chemical potential and energy gap values one can calculate the thermodynamic properties that we are interested in. In Sec. IV we discuss our conclusions and future work. Other thermodynamic properties as entropy and specific heat will be solved elsewhere.

\section{Boson-Fermion formalism}

The ideal BF ternary gas hamiltonian is 
\begin{equation}
H_{0}=\sum_{\mathbf{k}_{1},s_{1}}\epsilon _{\mathbf{k}_{1}}a_{\mathbf{k}%
_{1},s_{1}}^{\dagger }a_{\mathbf{k}_{1},s_{1}}^{{}}+\sum_{\mathbf{K}%
}E_{+}(K)b_{\mathbf{K}}^{\dagger }b_{\mathbf{K}}^{{}}-\sum_{\mathbf{K}%
}E_{-}(K)c_{\mathbf{K}}^{\dagger }c_{\mathbf{K}}^{{}} \label{H0}
\end{equation}%
where $\mathbf{K}\equiv \mathbf{k}_{1}+\mathbf{k}_{2}$ is the center-of-mass momentum (c.m.m.) wavevector of two fermions, $K \equiv \vert \mathbf{K}\vert $ and $\mathbf{k}\equiv \frac{1}{2}(\mathbf{k}_{1}-\mathbf{k}_{2})$ is their relative wavevector. The first term refers to the unbound fermions with energy $\epsilon _{k_{1}}\equiv \hbar ^{2}k_{1}^{2}/2m$, the second term refers to the bosonic 2eCPs while the third term is for bosonic 2hCPs both with $E_{\pm }(K)=E_{\pm}(0)+\hbar ^{2}K^{2}/4m$, respectively, with $E_{\pm }(0)$ for $K=0$.

The interaction hamiltonian $H_{int}$ has four BF interaction vertices, one with two-fermion/one-boson creation-annihilation operators representing how the unbound electrons (subindex $+$) or holes (subindex $-$) combine to form 2e/2hCPs in any $d=3,2,1$-dimensional system of size $L$. Thus in 3D 
\begin{eqnarray}
H_{int} &=& L^{-3/2}\sum_{\mathbf{k},\mathbf{K}} f_{+}(k)
\notag \\
&\times & \left( a_{\mathbf{k}+\frac{1}{2}\mathbf{K},\uparrow }^{\dagger }a_{-\mathbf{k}+\frac{1}{2}\mathbf{K},\downarrow }^{\dagger }b_{\mathbf{K}}+a_{-\mathbf{k}+\frac{1}{2}\mathbf{K},\downarrow }^{{}}a_{\mathbf{k}+\frac{1}{2}\mathbf{K},\uparrow }^{{}}b_{\mathbf{K}}^{\dagger } \right)   \notag \\
&+& L^{-3/2}\sum_{\mathbf{k},\mathbf{K}} f_{-}(k) 
\notag \\
&\times & \left( a_{\mathbf{k}+\frac{1}{2}%
\mathbf{K},\uparrow }^{\dagger }a_{-\mathbf{k}+\frac{1}{2}\mathbf{K}%
,\downarrow }^{\dagger }c_{\mathbf{K}}^{\dagger }+a_{-\mathbf{k}+\frac{1}{2}%
\mathbf{K},\downarrow }^{{}}a_{\mathbf{k}+\frac{1}{2}\mathbf{K},\uparrow
}^{{}}c_{\mathbf{K}}^{{}} \right)  \label{Hint}
\end{eqnarray}%
%
where $f_{\pm }(k)$ are the BF interaction functions defined in Refs.\cite%
{GBEC1, GBEC2} for electrons/holes. 

Neglecting nonzero $K$ values and ignoring those bosons with $K\neq 0$ in $H_{int}$---but \textit{not} in $H_{0}$ as assumed in BCS theory---one can consider a simpler \textit{reduced} $H_{red}$. Applying the Bogoliubov recipe of replacing the zero-$K$ creation operators $b_{\mathbf{0}}^{\dagger}$ and $c_{\mathbf{0}}^{\dagger}$ for the 2e/2hCP bosons by the numbers $\sqrt{N_{0}}$ and $\sqrt{M_{0}}$, with $N_{0}$ and $M_{0}$ the numbers of 2e/2hCP $K=0 $ bosons and applying the Bogoliubov-Valatin transformation \cite{bogo58, valatin} allows exactly diagonalizing\cite{GBEC3} the dynamical operator $\hat{H}_{red}-\mu \hat{N}$ with $\hat{N}$ the total-electron-number operator and $\mu $ a Lagrange multiplier. 

The thermodynamic (or Landau) potential of the grand-canonical statistical ensemble is now $\Omega (T,L^{3},\mu ,N_{0},M_{0})\break =-k_{B}T\ln \left[\mathrm{Tr}(\exp \{-\beta (\hat{H}_{red}-\mu \hat{N})\})\right] $ where \textrm{Tr} means \textquotedblleft trace,\textquotedblright\ $L^{3}$ is the
3D system volume and $\beta \equiv 1/k_{B}T$ with $k_{B}$ the Boltzmann constant and $\mu $ the electronic chemical potential. Specifically, the thermodynamic grand potential of this ternary BF mixture is
\begin{flalign} \label{omega}
\frac{\Omega}{L^{3}} &= \int\limits_{0}^{\infty }d\epsilon N(\epsilon_k )[\epsilon_k -\mu -E(\epsilon_k )]  \\ 
&- 2k_{B}T\int\limits_{0}^{\infty }d\epsilon N(\epsilon_k )\ln \left( 1+\exp [-\beta
E(\epsilon_k )]\right)  \notag \\
&+  [E_{+}(0)-2\mu ]n_{0} + [2\mu-E_{-}(0)]m_{0} \notag \\
&+ k_{B}T\int\limits_{0^{+}}^{\infty }d\varepsilon_K M(\varepsilon_K )\ln \left( 1-\exp
[-\beta (E_{+}(0)+\varepsilon_K -2\mu )]\right)  \notag \\
&+ k_{B}T\int\limits_{0^{+}}^{\infty }d\varepsilon_K M(\varepsilon_K )\ln \left( 1-\exp
[-\beta (2\mu -E_{-}(0)+\varepsilon_K )]\right) \notag 
\end{flalign}
where $n_0 \equiv N_0/L^3$, $m_0 \equiv M_0/L^3$ are the number densities of 2eCPs and 2hCPs with $K=0$, respectively, and $L^3$ the system volume in 3D. Here $N(\epsilon_k)$ is the fermionic density of states (DOS), $M(\varepsilon_K)$ the bosonic DOS and $E_{\pm}(0) = 2E_f \pm \delta\epsilon$ are the boson energies with $K=0$, with $E_f$ a pseudo-Fermi energy of the unbound fermions and $\delta\epsilon$ the BF interaction range energy to bound the CPs. Also, $E(\epsilon_k) = \sqrt{(\epsilon_k-\mu)^2 + \Delta^2(T)} $ with $\mu$ the BF chemical potential and $\Delta(T) = f_{+}\sqrt{n_0}(T) + f_{-}\sqrt{m_{0}(T)}$ the energy gap with $f_{\pm}$ the BF interaction functions defined as
\begin{equation*}
f_{+}(\epsilon_k ) \equiv \left\{ 
\begin{array}{c}
f\quad \text{if}\quad E_{f}<\epsilon_k <E_{f}+\delta \epsilon \\ 
0\quad \text{otherwise},%
\end{array}%
\right.  
\end{equation*}%
\begin{equation*}
f_{-}(\epsilon_k ) \equiv \left\{ 
\begin{array}{c}
f\quad \text{if}\quad E_{f}-\delta \epsilon <\epsilon_k <E_{f} \\ 
0\quad \text{otherwise}.%
\end{array}%
\right.  
\end{equation*}

From \eqref{omega} at $T=0$ one gets the ground state as
\begin{flalign}
\frac{\Omega_0}{L^3} &= [E_{+}(0)-2\mu_0 ]n_{0} + [2\mu_0-E_{-}(0)]m_{0}
\notag \\
&+ \int\limits_{0}^{\infty }d\epsilon N(\epsilon_k )[\epsilon_k -\mu_0 - \sqrt{(\epsilon_k-\mu_0)^2 + \Delta^2(0)}]
\end{flalign}
with $\Delta(0)= f_{+}\sqrt{n_0(0)} + f_{-}\sqrt{m_0(0)}$ the energy gap at $T=0$ and $\mu_0$ the chemical potential at $T=0$. The first two terms are the contribution of 2e/2hCPs with $K=0$ while the third term is the contribution of the unbound fermions interacting to bound 2e/2hCPs.

Imposing thermodynamic equilibrium conditions, i.e., minimizing the free energy with respect to the number of 2eCPs and 2hCPs, one gets
\begin{eqnarray}
\frac{\partial F}{\partial N_0} = 0 \hspace*{1.5cm} \frac{\partial F}{\partial M_0} = 0 \hspace*{1.5cm} \frac{\partial \Omega}{\partial \mu} = -N .\label{conditions}
\end{eqnarray}
From the first equation one gets the gap-like equation for 2eCPs
\begin{flalign}
&2\sqrt{n_0}[E_{+}(0) -2\mu] = \notag \\
&\int\limits_{E_{f}}^{E_{f}+\delta \epsilon} d\epsilon\;N(\epsilon_k )\; \frac{f_{+}\; \Delta_{e}(T)}{E(\epsilon_{k})} \tanh \left[ \tfrac{1}{2}\beta E(\epsilon_{k})\right] \label{gap-2e}
\end{flalign}
and from the second equation one gets the gap-like equation for 2hCPs
\begin{flalign}
&2\sqrt{m_0}[-E_{-}(0) +2\mu] = \notag \\
&\int\limits_{E_{f}-\delta \epsilon}^{E_{f}} d\epsilon\;N(\epsilon_k )\;
\frac{f_{-}\; \Delta_{h}(T)}{E(\epsilon_k)} 
\tanh \left[ \tfrac{1}{2}\beta E(\epsilon_k) \right] \label{gap-2h}
\end{flalign}
where $\Delta_{e}(T) = f_{+}\sqrt{n_0(T)}$ is the energy gap for 2eCPs and $\Delta_{h}(T) = f_{-}\sqrt{m_0(T)}$ the energy gap for 2hCPs. Here we consider the special case when $f_{+} = f_{-} = f$.

The third equation in \eqref{conditions} gives the number equation
\begin{flalign}
n &= 2n_0(T) - 2m_0(T)  + 2n_{B+}(T)  - 2m_{B+}(T) + n_f(T) \notag \\
&= 2 n_0(T) - 2m_0(T) \notag \\
&+ 2\int\limits_{0}^{\infty} d\varepsilon_K M(\varepsilon_K) \left(\frac{1}{\exp[\beta \{E_{+}(0) + \varepsilon_K-2\mu)\}]-1} \right) \notag \\
&- 2\int\limits_{0}^{\infty} d\varepsilon_K M(\varepsilon_K) \left(\frac{1}{\exp[\beta \{E_{-}(0) + \varepsilon_K+2\mu)\}]-1} \right) \notag \\
&+ \int\limits_{0}^{\infty}d\epsilon N(\epsilon_k )\left( 1-\frac{%
\epsilon_k -\mu }{E(\epsilon_k)}\tanh \left[ \tfrac{1}{2}\beta E(\epsilon_k)%
\right] \right) \label{number}
\end{flalign}
where the subindex $B+$ refers to 2e/2hCPs bosons with $K\neq 0$. The last term in \eqref{number} refers  to the unbound fermions. Hereafter, we define $n_f(T=0) \equiv n_{f0}$ as the number density of the unbound fermions at $T=0$. The system equations \eqref{gap-2e}, \eqref{gap-2h} and \eqref{number} are named the extended crossover BCS-Bose equations \cite{chavez17a, chavez17b} as there is an \textit{explicit} energy-gap equation for 2hCPs. Since this BF ternary gas is in thermodynamic equilibrium, and taking the thermodynamic limit $N\to \infty$ as well as $L^3 \to \infty$ but $n \equiv N/L^{3} = const$, implies the condition that the chemical potential is the same for all the particles involved.

When solving simultaneously the set of equations \eqref{gap-2e}, \eqref{gap-2h} and \eqref{number} in the weak-coupling (BCS regime), the resulting superconducting chemical potential has these characteristics: it is not constant but depends on temperature; its magnitude is slightly less than the normal case for temperatures from zero to $T_c$; where it shows a kink as a jump in its first derivative with respect to temperature. These superconducting specific heat features are closely related to both the magnitude of the temperature-dependent superconducting gap and to the Fermi energy of the superconductor (SC). 

\section{Theoretical Results}

Here we calculate the temperature dependence of the energy gap $\Delta(T)$, the chemical potential $\mu(T)$, the thermodynamic potential $\Omega(T)$, the Helmholtz free energy $F(T)$ and the condensation energy $E_{cond}(T)$ for the ternary BF gas mixture. When taking the 50-50 case between 2eCPs and 2hCPs, i.e., $n_0(T)=m_0(T)$ and $n_{B+}(T) = m_{B+}(T)$ the gap equations \eqref{gap-2e} and \eqref{gap-2h} can be added together since the integrand is the same, giving
\begin{equation}
\delta\epsilon = \frac{f^{2}}{2}  \int\limits_{E_{f}-\delta\epsilon}^{E_{f}+\delta \epsilon} d\epsilon \frac{N(\epsilon_k )}{\sqrt{(\epsilon_k - \mu)^2 + \Delta^2}} 
\tanh \left( \frac{\sqrt{(\epsilon_k - \mu)^2 + \Delta^2}}{2 k_B T} \right). \label{gap-5050}
\end{equation}
In order to relate our results to BCS, we identify $\delta\epsilon = \hbar\omega_D$ the Debye energy of the lattice and $f^2/2\delta\epsilon = V $ where $V$ is the BCS model interaction strength. The number equation \eqref{number} then becomes
\begin{flalign}
&n = \int\limits_{0}^{\infty}d\epsilon N(\epsilon_k )\left( 1-\frac{%
\epsilon_k -\mu }{\sqrt{(\epsilon_k - \mu)^2 + \Delta^2}}\tanh \left[ \frac{\sqrt{(\epsilon_k - \mu)^2 + \Delta^2}}{2 k_B T} \right] \right) \notag \\
 \label{number-5050}
\end{flalign}
\noindent since the integrands of $2n_{B+}(T)$ and $2m_{B+}(T)$ are equal. Both \eqref{gap-5050} and \eqref{number-5050} must be solved simultaneously to find the energy-gap and chemical-potential values. In terms of the Fermi energy one can rewrite the BF interaction constant $f$ as
\begin{equation*}
\tilde{G} \equiv \frac{f^2 m^{3/2}}{2^{5/2}\pi^{2}\hbar^{3}E_F^{1/2}} \geq 0
\end{equation*}
which is now the BF dimensionless \textit{strength} interaction. If we assume an electron-phonon interaction this strength interaction is related with the dimensionless coupling parameter of BCS theory $\lambda_{BCS}$ through $\tilde{G} = \lambda_{BCS}/2\delta\tilde{\epsilon}$ where tilde means dimensionless with respect to Fermi energy.

\begin{widetext}
The Helmholtz free energy $F(T,L^{3},\mu,N_{0},M_{0}) \equiv \Omega(T,L^{3},\mu,N_{0},M_{0}) + N \mu$  of this ternary BF gas mixture is 
\begin{eqnarray}
\frac{F}{L^{3}} &=&  \int\limits_{0}^{\infty }d\epsilon N(\epsilon_k )[\epsilon_k -\mu -E(\epsilon_k )] 
- 2k_{B}T\int\limits_{0}^{\infty }d\epsilon N(\epsilon_k )\ln \left( 1+\exp [-\beta
E(\epsilon_k )]\right)
\notag \\
&+& [E_{+}(0)-2\mu ]n_{0}
+ k_{B}T\int\limits_{0^{+}}^{\infty }d\varepsilon_K M(\varepsilon_K ) 
\ln \left( 1-\exp [-\beta (E_{+}(0)+\varepsilon_K -2\mu )]\right)
\notag \\
&+& [2\mu-E_{-}(0)]m_{0}
+ k_{B}T\int\limits_{0^{+}}^{\infty }d\varepsilon_K M(\varepsilon_K ) 
\ln \left( 1-\exp
[-\beta (2\mu -E_{-}(0)+\varepsilon_K )]\right)
+ n\;\mu(T).  \label{FreeH}
\end{eqnarray}

To calculate the SC Helmholtz free energy we must substitute the energy gap and chemical potential values in \eqref{FreeH}, and the normal Helmholtz free energy is calculated by setting $\Delta(T)= f\sqrt{n_0(T)} + f \sqrt{m_0(T)} =0$, i.e, there are no condensed 2e/2hCPs. Thus, one is able to get the chemical potential of the normal state (IFG). Explicitly, for the superconducting state we get

\begin{flalign}
\frac{F_{\mathtt{s}}}{L^{3}} &=
\bigg[ \int\limits_{0}^{E_f-\delta\epsilon} +  \int\limits_{E_f+\delta\epsilon}^{\infty} \bigg] \,d\epsilon_k N(\epsilon_k )[\epsilon_k -\mu_{\mathtt{s}} - |\epsilon_k - \mu_{\mathtt{s}}|]
%
+ \int\limits_{E_f-\delta\epsilon}^{E_f+\delta\epsilon} \,d\epsilon_k N(\epsilon_k )[\epsilon_k -\mu_{\mathtt{s}} -\sqrt{(\epsilon_k-\mu_{\mathtt{s}})^2 + \Delta^2}]
\notag \\
&- 2k_{B}T \bigg[ \int\limits_{0}^{E_f-\delta\epsilon} + \int\limits_{E_f+\delta\epsilon}^{\infty} \bigg] \,d\epsilon_k N(\epsilon_k )\ln \left( 1+\exp [-\beta
|\epsilon_k - \mu_{\mathtt{s}}|]\right)
%
%
- 2k_{B}T  \int\limits_{E_f-\delta\epsilon}^{E_f+\delta\epsilon} \,d\epsilon_k N(\epsilon_k )\ln \left( 1+\exp [-\beta
\sqrt{(\epsilon_k-\mu_{\mathtt{s}})^2 + \Delta^2}]\right)
\notag \\
&+ [2E_f + \delta\epsilon -2\mu_{\mathtt{s}} ]n_{0}
+ k_{B}T\int\limits_{0^{+}}^{\infty }d\varepsilon_K M(\varepsilon_K ) \ln \left( 1-\exp
[-\beta (2E_f + \delta\epsilon+\varepsilon_K -2\mu_{\mathtt{s}} )]\right)
\notag \\
&+ [-2E_f + \delta\epsilon + 2\mu_{\mathtt{s}}]m_{0}
+ k_{B}T \int\limits_{0^{+}}^{\infty }d\varepsilon_K M(\varepsilon_K ) \ln \left( 1-\exp
[-\beta (-2E_f + \delta\epsilon +\varepsilon_K + 2\mu_{\mathtt{s}} )]\right)
%
+ n\; \mu_{\textbf{s}}(T).  \label{FreeH5050}
\end{flalign}
\end{widetext}

Taking $\Delta(T) = 0$ one gets
\begin{flalign}
\frac{F_{\mathtt{n}}}{L^{3}} &=
\int\limits_{0}^{\infty} \,d\epsilon_k N(\epsilon_k )[\epsilon_k -\mu_{\mathtt{n}} - |\epsilon_k - \mu_{\mathtt{n}}|] \notag \\
&- 2k_{B}T  \int\limits_{E_f}^{E_f+\delta\epsilon} \,d\epsilon_k N(\epsilon_k )\ln \left( 1+\exp [-\beta
|\epsilon_k - \mu_{\mathtt{n}}|]\right)  \notag \\
&+ n\;\mu_{\mathtt{n}}(T).  \label{FreeHNormal}
\end{flalign}
Thus, the condensation energy is just
\begin{equation}
E_{cond}(T,n) = F_{\mathtt{s}}(T,n) - F_{\mathtt{n}}(T,n). \label{eq-Econd}
\end{equation}

\subsection{Chemical potential}
\begin{figure}[!ht]
	\centering
	\subfigure[]{\includegraphics[width=8.3cm]{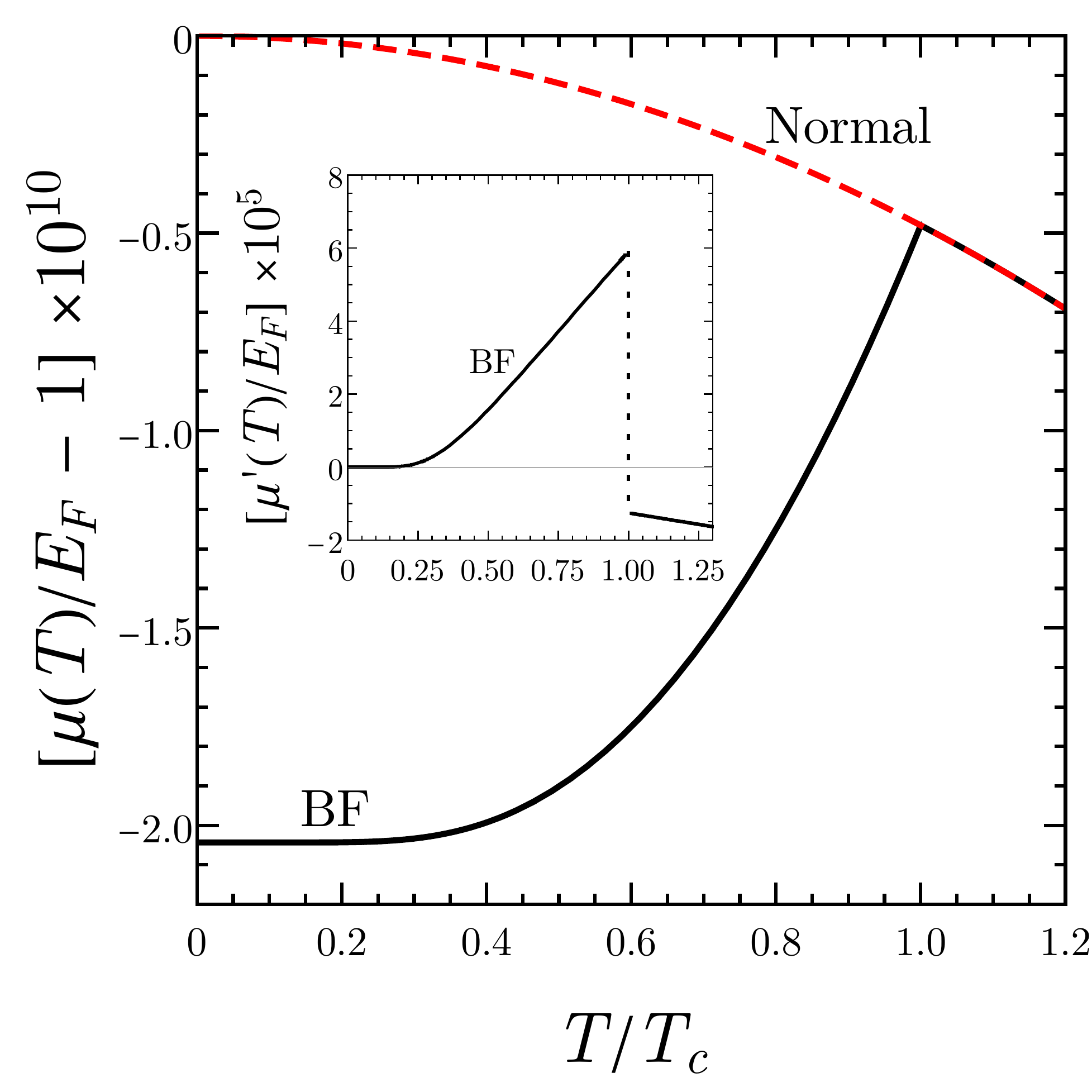}}
	\qquad
	\subfigure[]{\includegraphics[width=8.3cm]{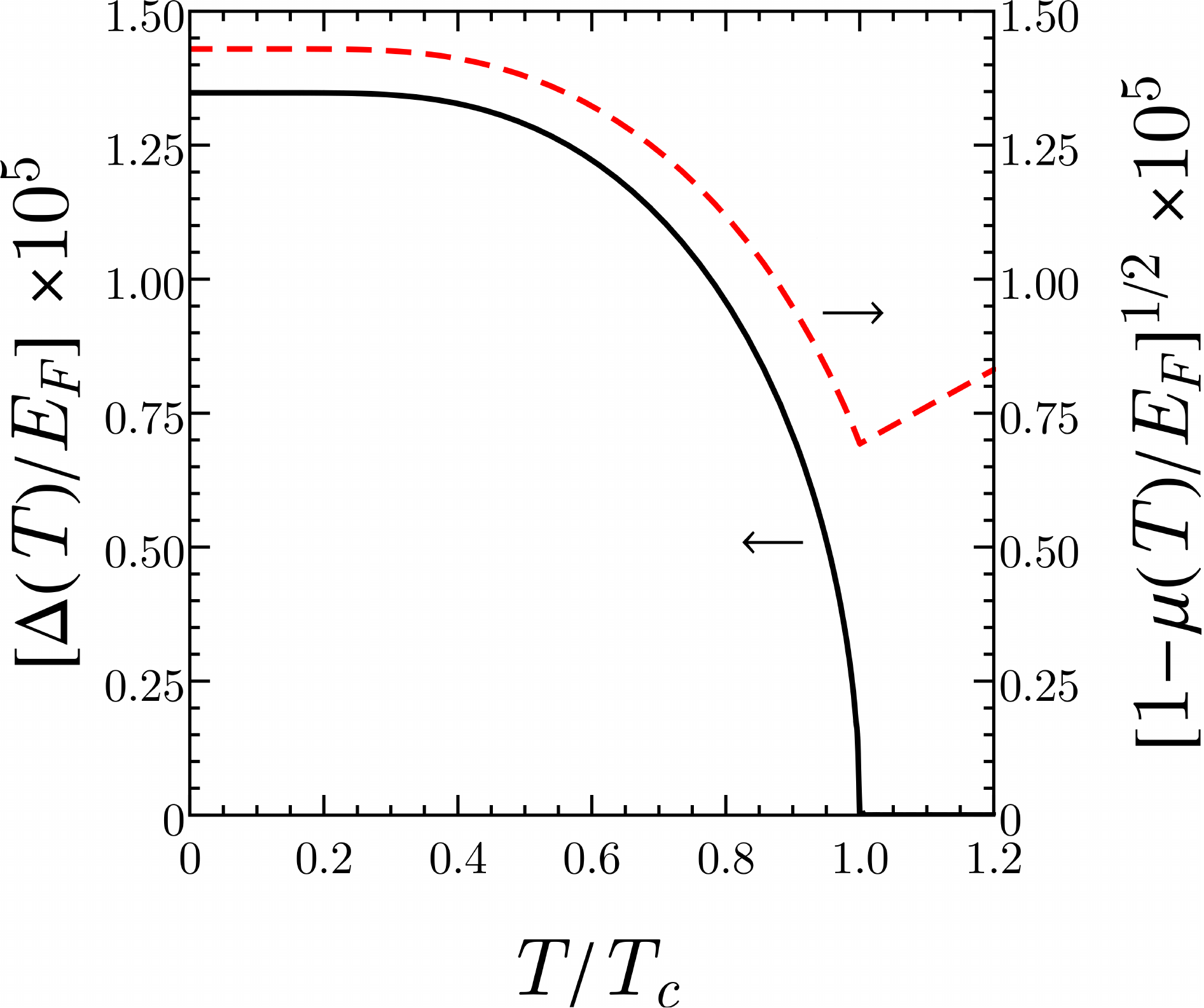}}
	\caption{(Color online) \textbf{a}) Chemical potential difference $\mu(T)/E_F -1$ vs. $T/T_c$ for BF (black full) curve alongside with normal (red dashed) curve. The SC chemical potential becomes equal to the normal state one precisely at $T_c$ after which both curves coincide. Inset shows the first derivative of the chemical potential $\mu^{\prime}(T)/E_F$ vs. $T/T_c$. Precisely at $T_c$ one sees the discontinuity cited in Sec. IIIA.
	\textbf{b}) Energy gap $\Delta(T)/E_F$ (black full) curve and the square root of the difference between the SC chemical potential and the normal chemical potential $[1-\mu(T)/E_F]^{1/2}$ vs. $T/T_c$ (red dashed) curve both in terms of the Fermi energy. We used in both figures $\delta\tilde{\epsilon} = 10^{-3}$ and $\tilde{G} = 10^{-4}$.}\label{ChemPot1}
\end{figure}

By solving \eqref{gap-5050} and \eqref{number-5050} simultaneously we plot in Fig.~\ref{ChemPot1}a the difference between the SC chemical potential and the normal (IFG) chemical potential at  $T=0$, i.e., $[\mu(T)/E_F-1]$. Note that there exists a finite difference between the superconductor (SC) and the normal (N) state of the order of $\simeq 2 \times 10 ^{-10} E_F$ for $T < T_c$. This quantity is usually omitted in BCS theory inasmuch as it was assumed that the SC chemical potential equals the Fermi energy at zero absolute temperature. At $T=T_c$ there is a kink in $\mu_{\mathtt{s}}$ and thereafter the SC and normal chemical potentials coincide, as expected, meaning that interactions between fermions no longer form CPs so that the ternary BF mixture becomes an ideal Fermi gas. Although $\mu(T)$ is a continuous function, the first derivative of the chemical potential has a discontinuity at $T_c$, as shown in the Inset of Fig.~\ref{ChemPot1}a. The kink in the chemical potential and the discontinuity in its first derivative confirm the existence of a phase transition as the superconducting BF mixture changes to only unbound fermions, i.e., a normal state.

In Fig.\,\ref{ChemPot1}b we plot the energy gap $\Delta(T)/E_F$ along with $[1-\mu(T)/E_F]^{1/2}$, i.e.,  the square root of the difference between the SC chemical potential of the BF mixture and the normal state, where the two quantities are of the same order of magnitude and proportional to $(\Delta(0)/E_F)^2$ \cite{vandermarel}. From the BF model we obtain a factor equal to $0.94$ while in Ref.~\cite{vandermarel} a factor of $\sqrt{2}$ is reported. The energy gap has the familiar half-bell shape while the chemical potential has the kink that van der Marel \cite{vandermarel} pointed out, i.e., precisely at $T=T_c$. Also, both curves  have the same order of magnitude at $T=0$ but not the same value. This can be seen from the fact that we have a BF mixture with unbound electrons/holes to form 2e/2hCPs and that $\mu(T)$ gives the energy of the system to add/extract particles.

\subsection{Condensation energy}

To find the SC condensation energy one must first calculate the superconducting thermodynamic potential $\Omega_{\mathtt{s}}(T)$ along with the corresponding normal state one. Once we know the chemical potential $\mu_{\mathtt{s}}(T)$ and the energy gap $\Delta(T)$ values we can substitute them in \eqref{omega}. 

\begin{figure*}[!ht]
\centering
\subfigure[]{\includegraphics[width=8.6cm]{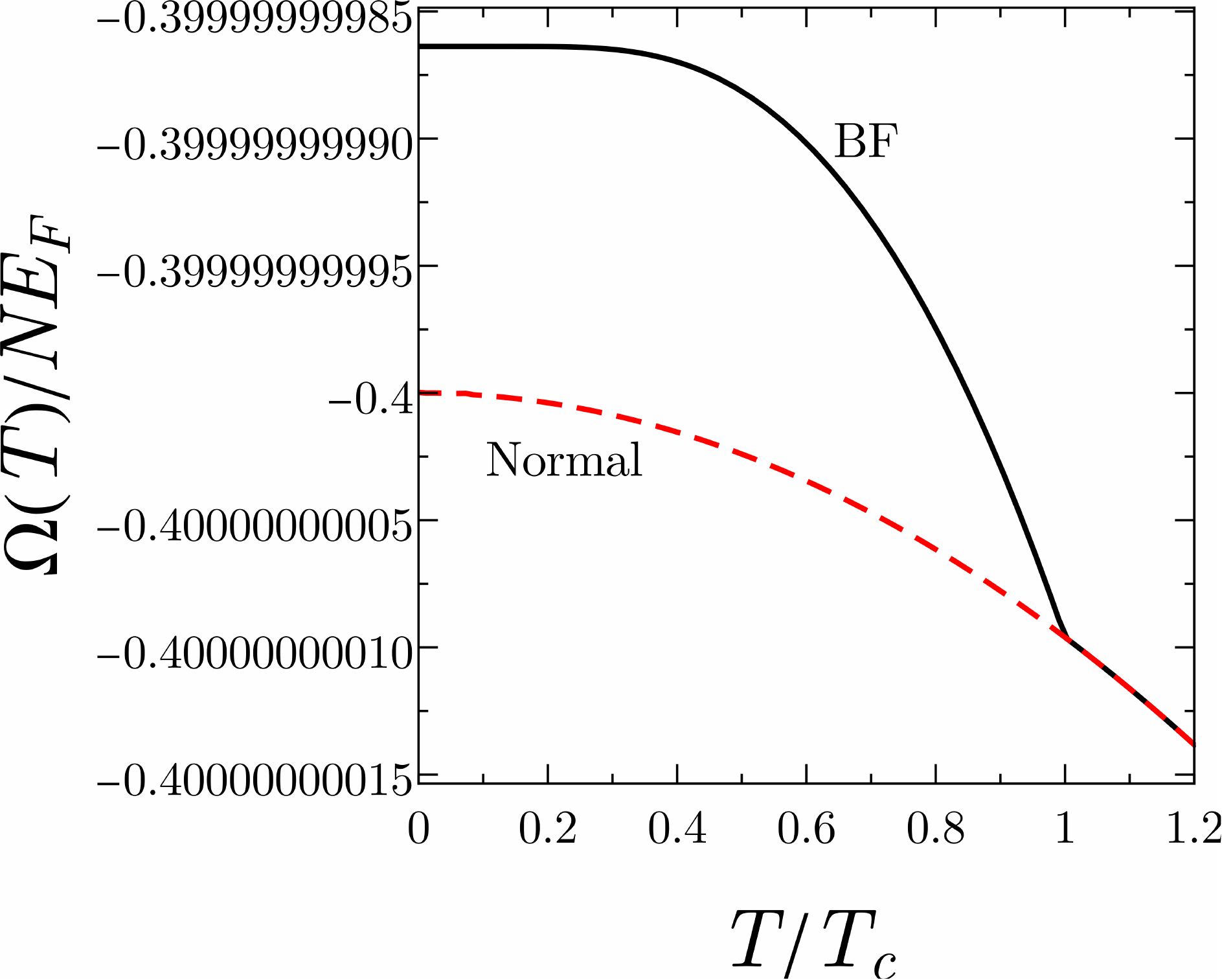}}
	\qquad
	\subfigure[]{\includegraphics[width=8.0cm]{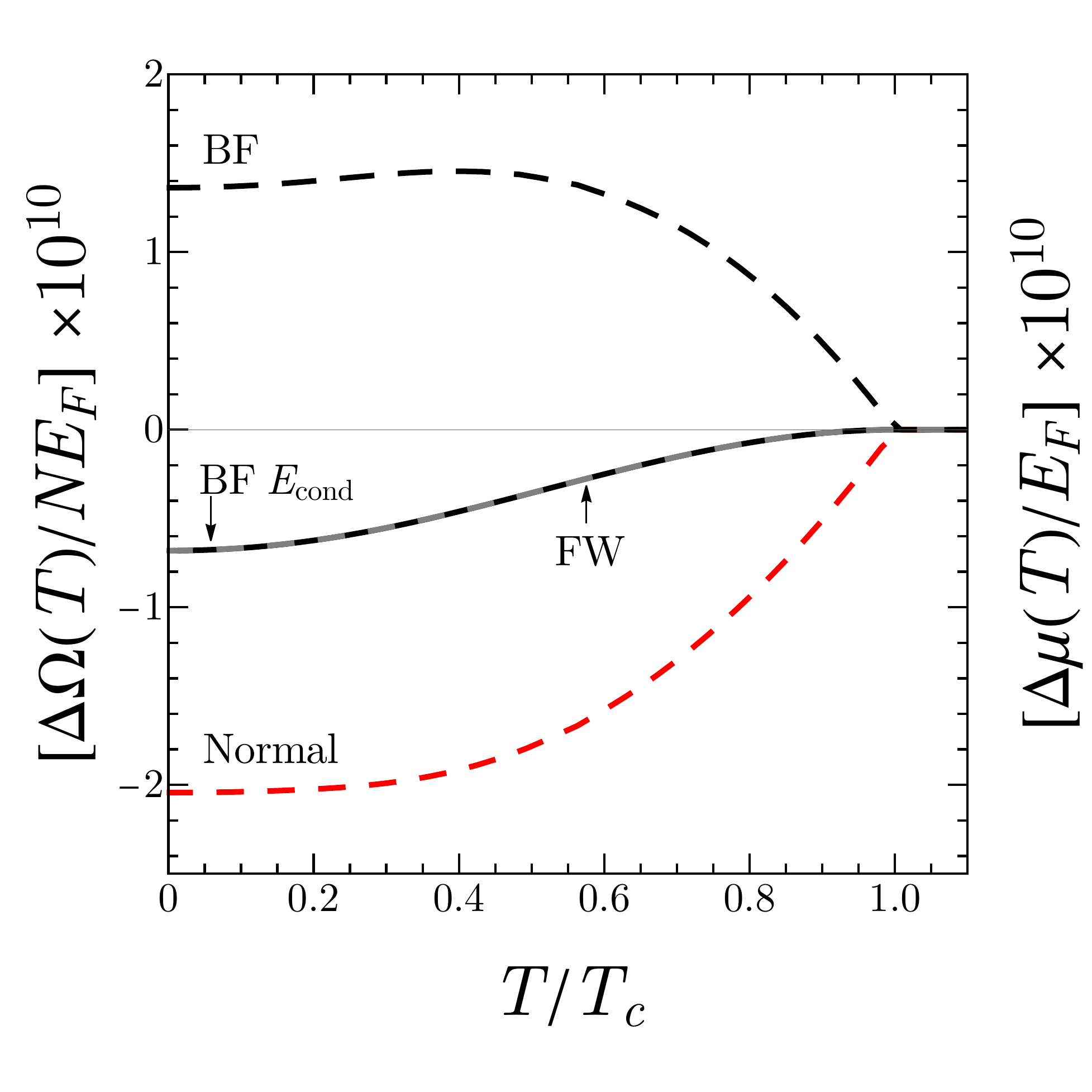}}
		\caption{(Color online) \textbf{a}) Thermodynamic potential $\Omega(T,n)/NE_F$ of BF SC state (black full) curve and normal state (red dashed) curve vs. $T/T_c$. The SC and normal curves coincide at $T=T_c$, and after that both systems show the same behavior.
		\textbf{b}) The difference of the thermodynamic potential between SC and normal states $\Delta \Omega = [\Omega_{\mathtt{s}}(T) - \Omega_{\mathtt{n}}(T)]/NE_F$ (black dashed) curve is shown together with the difference between SC and normal chemical potential $\Delta \mu = [\mu_{\mathtt{s}}(T)- \mu_{\mathtt{n}}(T)]/E_F$ (red dashed) curve. Also shown is the $\Delta \Omega$ (gray dashed) curve using Eq. (51.53) of Ref.~\cite{fetter} and $E_{cond}(T)$ the BF condensation energy (black full) curve. Four curves coalesce for $T \geq T_c$ becoming zero.}\label{Fig-Omega}
\end{figure*}

In Fig.~\ref{Fig-Omega}a we plot the SC thermodynamic potential \eqref{omega}, which corresponds to the BF mixture, together with the normal state. The SC thermodynamic potential is slightly greater than its normal state counterpart, but at $T=T_c$ both curves coincide, as expected, and remain equal for all $T \geq T_c$. In Fig. \ref{Fig-Omega}b we plot the difference between the SC and normal thermodynamic potential $\Delta\Omega = [\Omega_{\mathtt{s}}(T)- \Omega_{\mathtt{n}}(T)]/NE_F$ along with the difference between the SC and normal chemical potentials $\Delta\mu = [\mu_{\mathtt{s}}(T)- \mu_{\mathtt{n}}(T)]/E_F$ and Eq. (51.53) of Ref.~\cite{fetter}, i.e., the difference between SC and normal thermodynamic potentials only. The BF thermodynamic potential difference $\Delta\Omega_{BF}$ and chemical potential difference $\Delta\mu_{BF}$ are of the same order of magnitude and both are essential to correctly obtain the condensation energy, while the FW \cite{fetter} thermodynamic potential difference is an order of magnitude lower than the BF one. We find that $\Delta\mu_{BF} <0$ while $\Delta\Omega_{BF} >0$, both will be crucial to obtain the accurate condensation energy \cite{kim} from the difference between superconductor and normal Helmholtz free energy, while the FW \cite{fetter} difference is simply $\Delta\Omega_{FW} <0$. Furthermore, we found that $\mid \Delta\Omega_{BF} \mid~ \lesssim~ \mid \Delta \mu_{BF}\mid$ which is not addressed in the literature.

\begin{figure}[!ht]
\centering
\includegraphics[width=8.25cm]{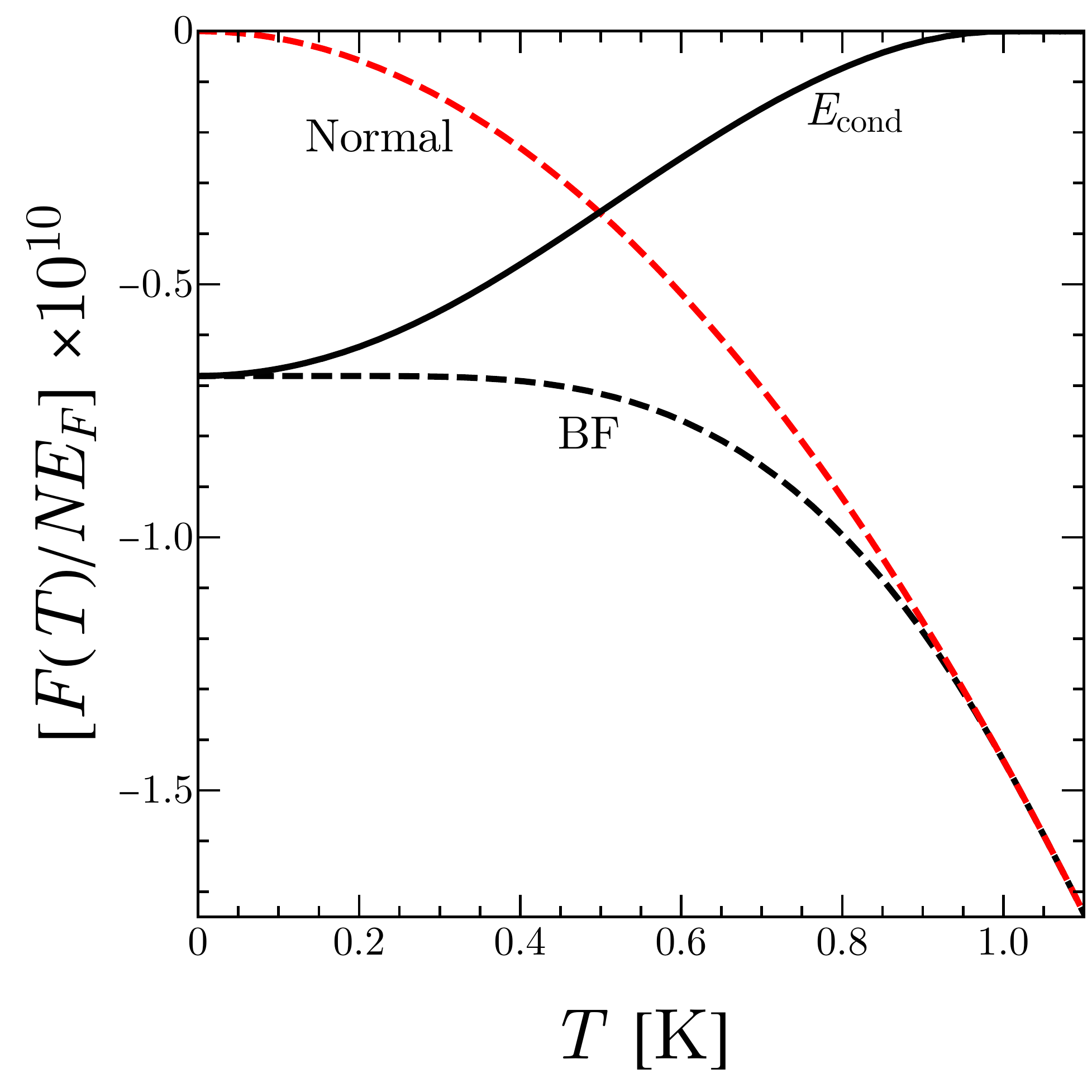}
\caption{(Color online) Helmholtz free energy $F_{\mathtt{s}}(T,n)$ for the BF mixture (black) curve and $F_{\mathtt{n}}(T,n)$ for the normal state (red) curve solving \eqref{FreeH5050} and \eqref{FreeHNormal}, respectively. Superconducting free energy curve becomes the normal state curve after $T_c$, as expected. Note that the $F_{\mathtt{s}}(T=0) - F_{\mathtt{n}}(T=0)$ already predicts the condensation energy at $T=0$. Also shown is the condensation energy for the BF mixture using \eqref{eq-Econd} (black dashed) curve. Here $\tilde{G} = 10^{-4}$ and $\delta\tilde{\epsilon} = 10^{-3}$ were used.}
\label{Fig-Helmholtz}
\end{figure}

\begin{table*}[!htbp]
\begin{center}
\caption{Reported data of the critical temperature, energy gap  and the condensation energy and the difference of chemical potential with Fermi energy at zero absolute temperature for some elemental SCs compared with that from numerical solutions by solving \eqref{gap-5050} and \eqref{number-5050} to find $T_c$ and $\Delta(T=0)$. The condensation energy $E_{cond}(T=0)$ for the same SC was obtained by substituting in \eqref{eq-Econd} the energy gap and chemical potential values. The last two columns list the BF parameters $\delta\epsilon$ and $\tilde{G}$ used for each SC. BCS critical temperatures, energy gap and condensation energy values are also included.}\label{Table1}
\setlength{\tabcolsep}{4pt}
\renewcommand{\arraystretch}{1.2}
\begin{tabular}{c|ccc|ccc|ccc|c|cc}
\toprule
 & \multicolumn{3}{c|}{$T_c$ [K]}
 & \multicolumn{3}{c|}{$\Delta(0)$ [meV]}
 & \multicolumn{3}{c|}{$E_{cond}(T=0)$ [mJ/mol]}
 & \multicolumn{1}{c|}{$1-\mu(0)/E_F$}
 & \multicolumn{2}{c}{BF parameters} \\
 & Exptl
 & BCS$^{\star}$ 
 & BF
 & Exptl
 & BCS$^{\star \star}$ 
 & BF 
 & Exptl 
 & BCS$^{\star \star \star}$ 
 & BF 
 & $\times 10^{-8}$
 & $\delta\tilde{\epsilon} \times 10^{-4}$ 
 & $\tilde{G}  \times 10^{-5}$ \\
\midrule
Al  & 1.17 & 1.16 & 1.17  & 0.16  & 0.176 & 0.174 & -0.42 & -0.38 & -0.39 & 0.03 & 30   & 25   \\
Tl  & 2.38  & 2.31 & 2.33 & 0.39  & 0.35  & 0.35  & -2.5  & -2.17  & -2.2  & 0.15  &  8.2  & 11.3    \\
In  & 3.46 & 3.43 & 3.46  & 0.52  & 0.59  & 0.52  & -5.0   & -6.7 & -4.6   & 0.28 & 11    & 15   \\
Sn  & 3.73 & 3.9  & 3.78  & 0.55  & 0.52  & 0.57 & -6.26   & -4.6 & -6.15   & 0.30 & 16    & 20  \\ 
Pb  & 7.19 & 7.27 & 7.27  & 1.34  & 1.11  & 1.11  & -46.8  & -34.5 & -35.5 & 0.76 & 8.7 & 16  \\ 
Nb  & 9.2 & 9.09 & 9.14  & 1.52 & 1.40  & 1.38 & -164.6  & -15.7 & -17.4  & 5.14 & 4.4   & 62   \\
\bottomrule
\end{tabular}
\end{center}
\begin{tablenotes}
{\small
\item $^{\star}$ Calculated with Eq. (3.29) from Ref. \cite{BCS}; $^{\star \star}$ Calculated with Eq. (2.40) from Ref. \cite{BCS}; $^{\star \star \star}$ Calculated with Eq. (3.38) from Ref.\cite{BCS} when $T \to 0$. Al data taken from \cite{biondi}; Tl data taken from \cite{keesom}; In data taken from \cite{tinkham}; Sn, Pb and Nb data taken \cite{townsend}; Condensation energy data taken from \cite{kim}.
}
\end{tablenotes}
\end{table*}

Fig.~\ref{Fig-Helmholtz}a shows the plot of $F(T) = \Omega(T) + N\mu(T)$ the Helmholtz free energy for the BF mixture using \eqref{FreeH5050} and the normal state \eqref{FreeHNormal}, substituting the corresponding energy gap and chemical potential values for all temperatures. The difference between the SC free energy and the normal free energy at $T=0$ gives the condensation energy $E_{cond}(0)$ but here we calculate it for every $0\leq T \leq T_c$, and for $T \geq T_c$. Fig.\,\ref{Fig-Helmholtz}b shows the condensation energy for the BF mixture which according using \eqref{eq-Econd} can be rewritten as
\begin{eqnarray}
E_{cond}(T) &=& F_{\mathtt{s}}(T) - F_{\mathtt{n}}(T)
\notag \\
&=& [\Omega_{\mathtt{s}}(T) + N\mu_{\mathtt{s}}(T)] - [\Omega_{\mathtt{n}}(T) + N \mu_{\mathtt{n}}(T)]
\notag \\
&=& [\Omega_{\mathtt{s}}(T) - \Omega_{\mathtt{n}}(T)] + [N \mu_{\mathtt{s}}(T) - N \mu_{\mathtt{n}}(T)]
\notag \\
&=& \Delta \Omega + N\Delta \mu.  \label{Econd}
\end{eqnarray}
The first and second terms of \eqref{Econd} are plotted in Fig.~\ref{Fig-Omega}b using black full and red dashed curves, respectively, where we find that the magnitude of $\Delta \Omega$ is slightly less than $\Delta\mu$ but with opposite sign. This implies that the condensation energy \eqref{Econd} will have the correct sign as reported in Ref.~\cite{fetter}. If we ignore the difference between SC and normal chemical potentials as in Ref.~\cite{fetter} the condensation energy would be $\sim 10^{-11} N\,E_F$, this order of magnitude is less than the BF calculations given here as previously discussed. In order to compare with reported data the difference of chemical potential $\mu_{\mathtt{s}} - \mu_{\mathtt{n}}$ is crucial.

\begin{figure}[!ht]
	\centering
	\subfigure[]{\includegraphics[width=8.25cm]{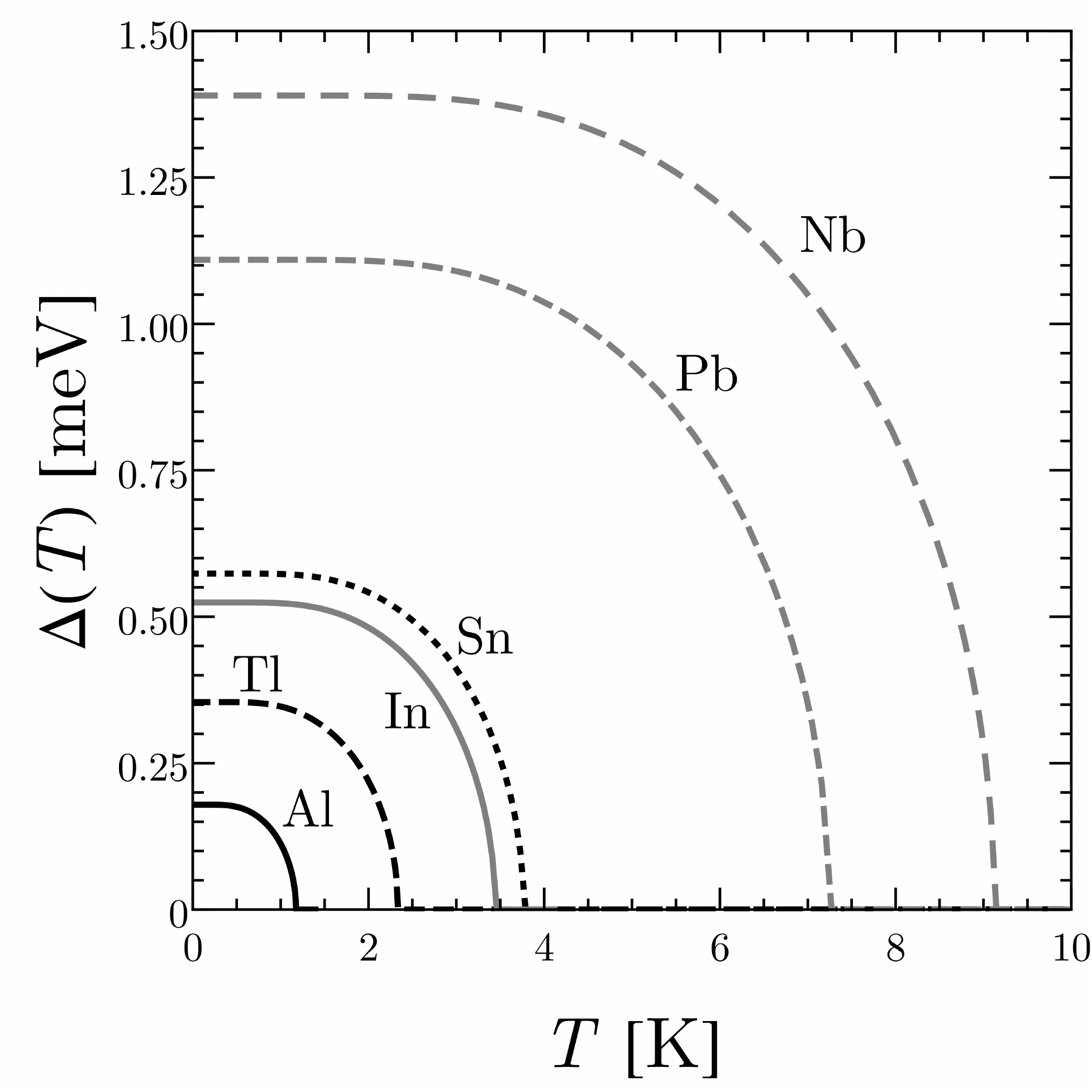}}
	\subfigure[]{\includegraphics[width=8.25cm]{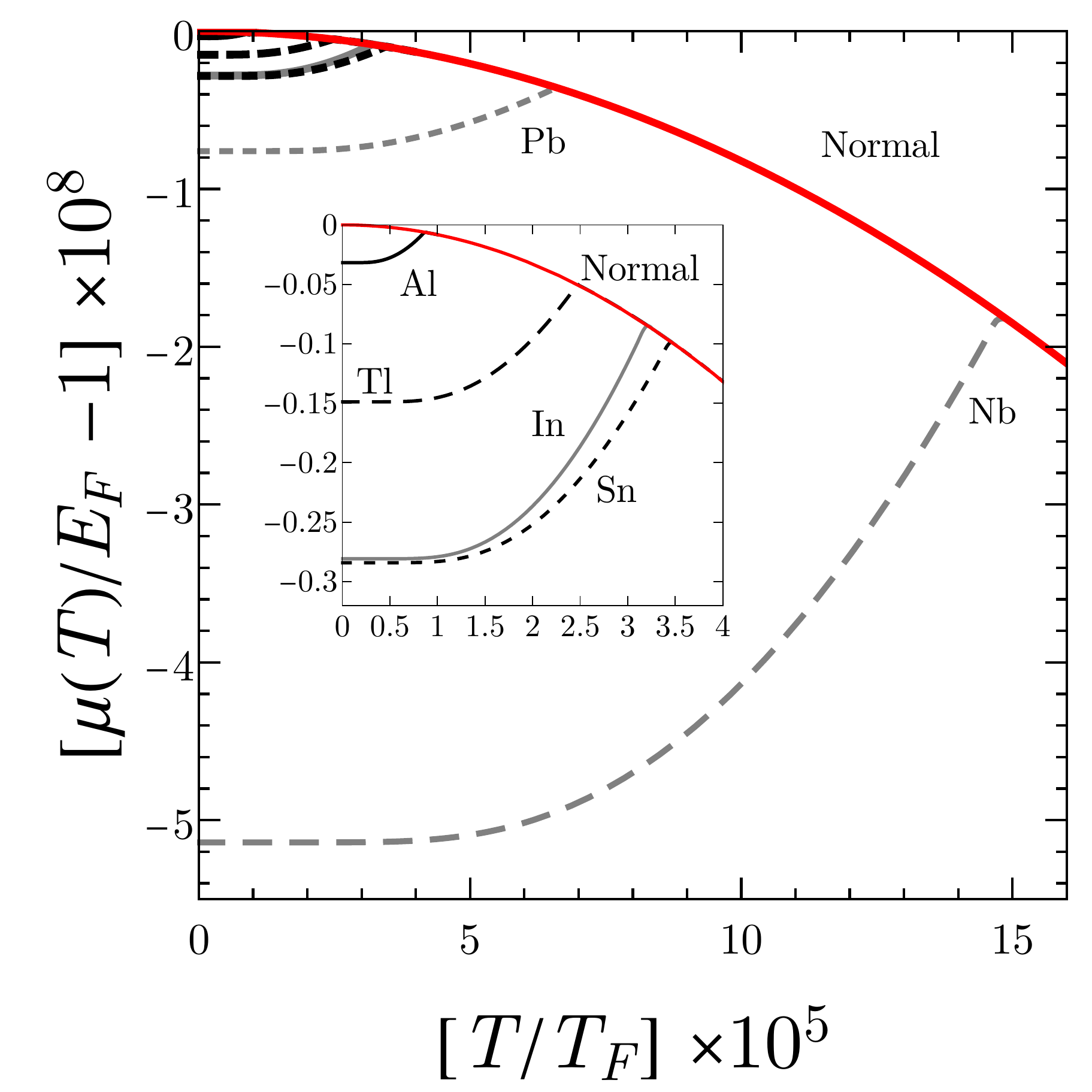}}
	\caption{(Color online) \textbf{a}) Energy gap $\Delta(T)$ [meV] vs. $T$ [K] for the SCs Al, Tl, In, Sn, Pb and Nb. All curves have the half-bell shape and become null precisely in $T_c$ for each SC. At $T=0$ the values are the ones listed in Table\,\ref{Table1}.
		\textbf{b}) Difference of chemical potential $\mu(T)/E_F - 1$ vs. $T/T_F$ for SCs Al, Tl, In, Sn, Pb and Nb using the BF model curves, compared with the normal (IFG) chemical potential (red full) curve. At precisely $T_c$, both SC and normal states become equal and beyond $T_c$ remain the same. The Inset shows the difference of the chemical potential for Al, Tl, Sn and In. Note the kink at precisely $T_c$.}
	\label{Fig-GapMu-elements}
\end{figure}

\subsection{Comparison with experimental data}
Here we compare our results with data for some elemental SCs in the weak-coupling extreme, i.e., with 50--50 proportions. Table \ref{Table1} shows the critical temperature $T_c$, the energy gap $\Delta(T=0)$ and the condensation energy $E_{cond}(T=0)$ for Al, Tl, In, Sn, Pb, and Nb using \eqref{gap-5050}, \eqref{number-5050} and \eqref{FreeH5050} with \eqref{FreeHNormal}, together with the corresponding experimental values. Also shown is the Debye energy $\delta\tilde{\epsilon}$ and the BF interaction strength $\tilde{G}$ for each SC. We notice that the ascending trend of the critical temperature values coincides with that of the energy gaps, as well as with those for the condensation energies, except for Pb and Nb.

For Pb we calculated $2\Delta(0)/k_BT_c \simeq 3.51$ (weak-coupling SC) but it is reported $\simeq 4.1$ \cite{tinkham} implying that Pb is a strong-coupling SC. Pb is known as a ``bad actor'' \cite{badactors} of the BCS theory. Although critical temperature is well reproduced here, energy gap and condensation energy are below of data trends. Here we addressed the weak-coupling extreme and is expected that Pb data cannot be reached. To do this, one must change the number density of unbound electrons which is related with the BF strength interaction; this will be reported elsewhere.

For Nb we obtained $2\Delta(0)/k_BT_c \simeq 3.51$ while the reported data goes from  $\simeq 3.66$ \cite{finnemore1} up to $\simeq 3.84$ \cite{townsend}, suggesting that Nb is not \cite{leupold64} a conventional SC but an intermediate one \cite{finnemore2}. Furthermore, Refs.~\cite{finnemore2,leupold64,kerchner} report two magnetic critical fields ($H_{c1}$, $H_{c2}$) which confirm that Nb is a type-II SC. An extension of this ternary BF model is required to appropriately describe a type-II SC, will be dealt with eventually.

The important difference here is that the BF results have been calculated using variable DOS and temperature-dependent chemical potentials (SC and normal), while BCS as well as FW \cite{fetter} results are calculated with DOS and chemical potentials taken as constants, both in SC and normal states. These results are discussed further in Ref.~\cite{chavez21b}.

Fig.~\ref{Fig-GapMu-elements}a shows the energy gap $\Delta(T)$ [meV] vs. $T$ [K] for the SCs Al, Tl, Sn, Pb and Nb, by solving \eqref{gap-5050} and \eqref{number-5050} simultaneously for the BF mixture, where the half-bell shape is preserved in all cases and the energy gap is zero at precisely $T_c$. The energy gap values at $T=0$ are the ones shown in Table\,\ref{Table1}. Fig.\,\ref{Fig-GapMu-elements}b shows the chemical potential difference $[\mu(T)/E_F -1]$ for the same SCs, and the kink appears for each one at the corresponding $T_c$; the general behavior after the critical temperature follows the IFG chemical potential.

\begin{figure}[!ht]
\centering
\includegraphics[width=8.25cm]{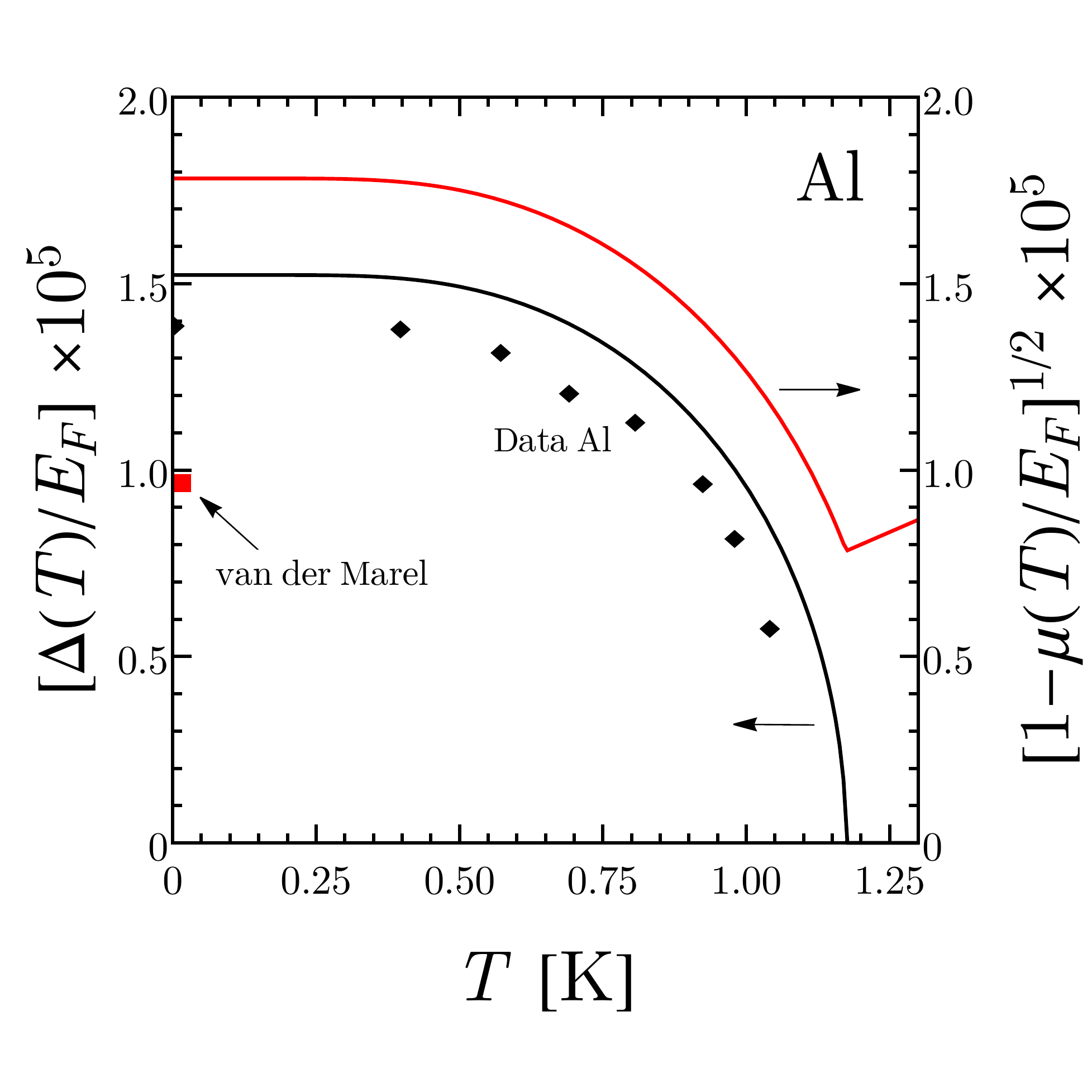}
\caption{(Color online) Energy gap $\Delta(T)/E_F$ (black full) curve and square root of chemical potential difference $[1 - \mu(T)/E_F]^{1/2}$ (red full) curve vs. $T/T_F$ for Al for the 50-50 case using $\delta\tilde{\epsilon}$ and $\tilde{G}$ from Table \ref{Table1}. The red square shows van der Marel's result applying \eqref{vandermarel} to the value $\Delta(T)/E_F$ at $T=0$. Also, the energy gap data \cite{biondi} is plotted.}\label{Fig-Mu-Gap-Al}
\end{figure}

\begin{figure}[!ht]
\centering
\subfigure[]{\includegraphics[width=8.6cm]{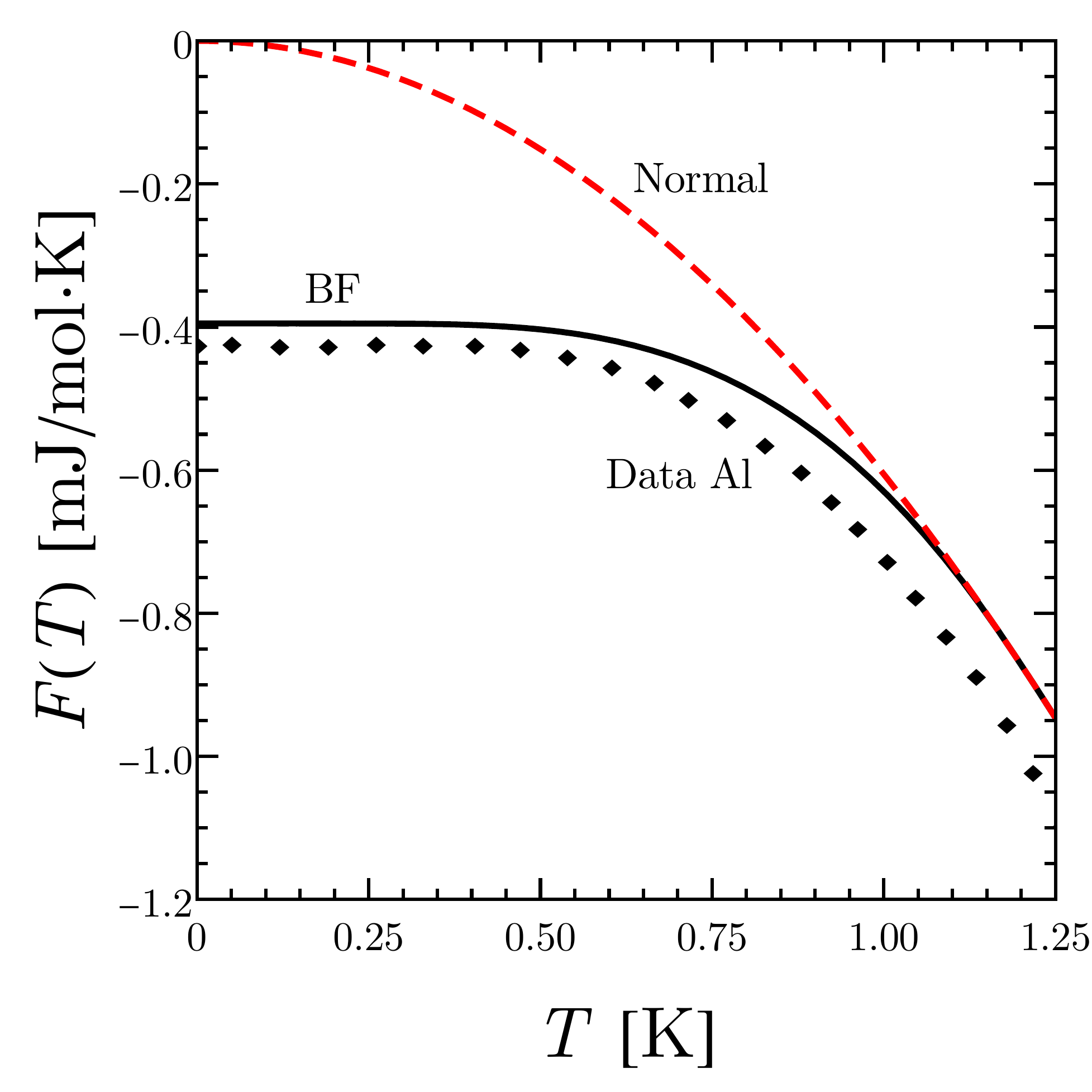}}
\qquad
\subfigure[]{\includegraphics[width=8.5cm]{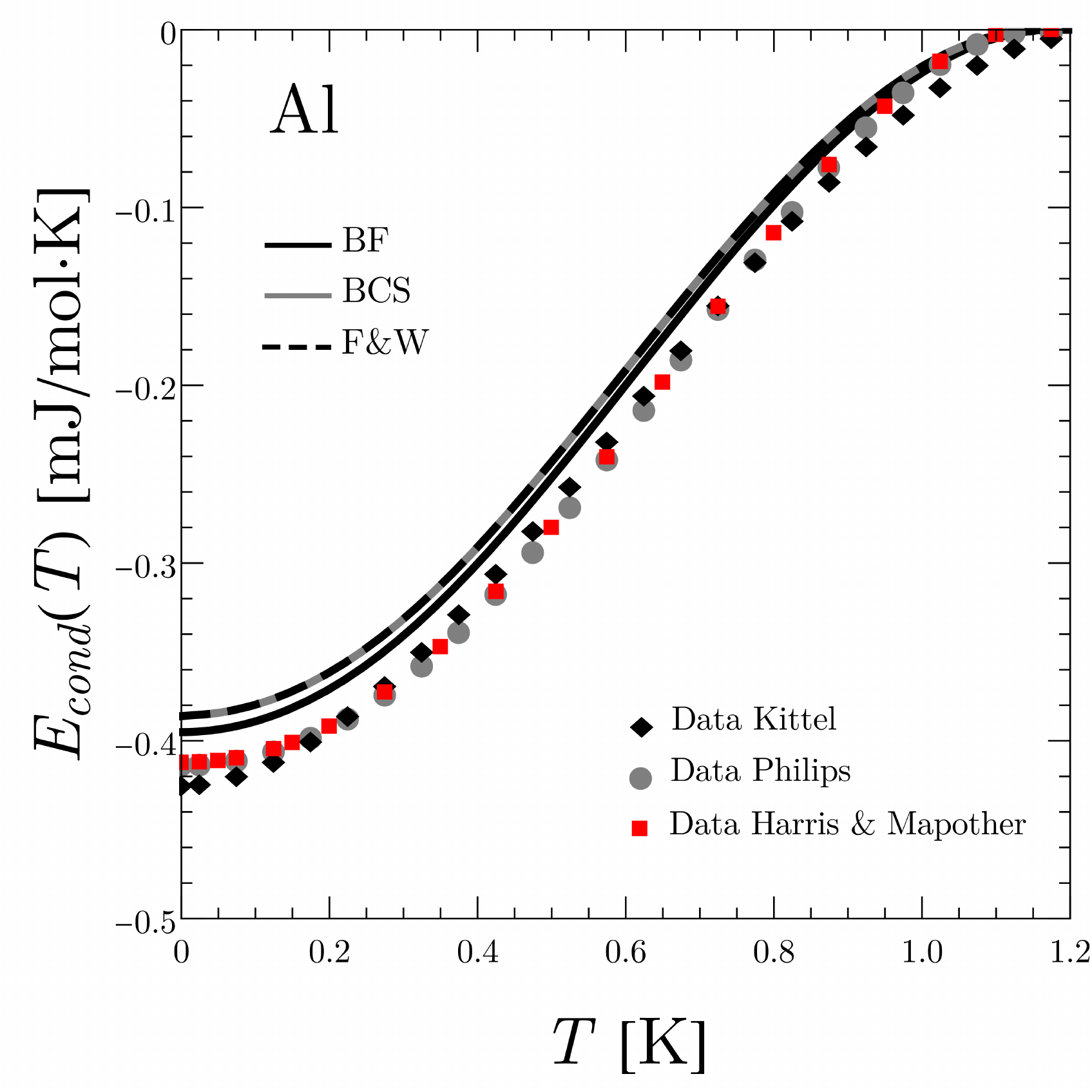}}
\caption{(Color online) \textbf{a}) Helmholtz free energy vs. temperature for 50-50 (black) curve and normal (red) curve cases compared with reported Al data \cite{kittel}. Note that the BF curve follows the behavior of data \cite{kittel}.
\textbf{b}) Condensation energy $E_{cond}(T)$ [mJ/mol·K] vs. $T$ [K] of BF mixture (black) curve, BCS (gray) curve, FW (black dashed) curve and Al data taken from Refs. \cite{kittel, philips}. }\label{Fig-EcondAl}
\end{figure}

Fig.~\ref{Fig-Mu-Gap-Al} shows the energy gap $\Delta(T)/E_F$ along with the square root of the difference of chemical potentials $[1 - \mu(T)/E_F]^{1/2}$ for Al. Once again, the energy gap and the difference of the chemical potentials have the same order of magnitude, the energy gap has the half-bell shape shown by the data, and the difference of chemical potentials follows the behavior of Fig.~\ref{ChemPot1}b exhibiting the kink previously mentioned. 

Fig.~\ref{Fig-EcondAl}a shows the Helmholtz free energy in the SC and normal states for Al, solving \eqref{FreeH5050} and \eqref{FreeHNormal} respectively, using the energy gap and chemical potential values as well as the BF parameters listed in Table \ref{Table1}. Also plotted are the Al reported data from Ref. \cite{kittel}; the BF mixture curve follows the behavior of the data even as it lies slightly above. This difference might be due to the fact that BF model does not consider interactions among composed bosons and unbound fermions, and bosons with $K \neq 0$ are neglected, which might manifest itself in thermodynamic properties whose order of magnitude is as small as in the condensation energy.

Fig.~\ref{Fig-EcondAl}b shows the condensation energy compared with Al reported data from Refs.~\cite{kittel,harris,philips}. For Al we used the $\delta\tilde{\epsilon}$ and $\tilde{G}$ values listed in Table~\ref{Table1}. Also shown are the BCS condensation energy curve plotted with Eq.\,(3.38) with BCS energy gap solutions Eq.\,(2.40) of  Ref.~\cite{BCS}, as well as the ones reported in Eq.\,(51.53) of Ref.~\cite{fetter}. The main difference between these three curves lies in that the DOS and the chemical potential were taken as constant in the BCS and FW solutions for both normal and SC states, while in the BF mixture they were taken as DOS variable and chemical potential temperature-dependent for both SC and normal states. Note that although other fluctuation \cite{vandermarel2002} contribute to the specific heat and the condensation energy, they are not taken into account here.

\section{Conclusions}

Using a Boson-Fermion model of superconductivity we calculated, among other properties, the chemical potential and the grand thermodynamic potential, which are necessary to obtain the Helmholtz free energy for both normal and superconducting states and from their difference, the condensation energy with 50-50 proportions between 2eCPs and 2hCPs in the weak-coupling regime. Commonly, to obtain the condensation energy, only the SC and normal thermodynamic potentials are added in order to obtain the Helmholtz free energy, ignoring the corresponding chemical potentials by arguing the smallness of the difference between the superconducting and normal chemical potentials. Here we did not make this assumption and found that the difference between SC and normal chemical potentials is slightly greater than the difference between SC and normal thermodynamic potentials, leading to the correct condensation energy value.

We show that the superconducting chemical potential differs from that assumed by the BCS theory, as it is temperature dependent and different from the Fermi energy at least within a $10^{-10} E_F$, which is of the same order as the condensation energy. We also show that $[1 - \mu_{\mathtt{s}}(T)/E_F]^{1/2} = \alpha~\Delta(T)/E_F$ not only for $T=0$ as van der Marel suggested for 2D superconductors with $\alpha = 1/\sqrt{2}$, which is extended here for 3D where $\alpha = 0.94$. Furthermore, we show the kink in the chemical potential, and the discontinuity in its first derivative at precisely $T_c$ confirming the existence of a phase transition.

The chemical potential has a remarkable influence on the superconducting thermodynamic properties, particularly on the condensation energy magnitude, as the magnitude of the difference between the superconducting and normal chemical potentials is of the same order as that for the thermodynamic potentials. For elemental SCs we highlight the inappropriateness of using the same superconducting and normal chemical potentials as well as leaving the DOS constant around the Fermi energy, which becomes more notorious as we consider high-temperature superconductors, as Fe-based and cuprates. Exact temperature dependent chemical potentials and variable DOS are essential for accurate condensation energy calculations, and therefore for critical fields.

Our condensation energy values reproduce quite well the reported experimental data for some SCs, except for Pb and Nb. For Al the results are slightly above from experimental data since those condensation energy values are obtained from the specific heat or from the critical thermodynamic field using the BCS formalism with its corresponding approximations \cite{leupold64}. Among other results, this confirms the goodness of the BF formalism in the weak-coupling limit when the number of pairs of particles is equal to the number of pairs of holes. Extensions to the present work including strong-coupling as well as asymmetry between the number of particle-pairs and the number of hole-pairs is in progress.

\subsection*{Acknowledgments}
IC thanks CONACyT (Mexico) for the Postdoc grant EPA1 \#~869450. PS, OAR and MAS thank PAPIIT-DGAPA-UNAM (Mexico) for grant IN110319. MdeLl thanks PAPIIT-DGAPA-UNAM (Mexico) for grant IN115120.

\balance

\end{document}